\documentclass[12pt,nohyper,letterpaper]{JHEP3}         % For preprint
\usepackage{amssymb,amsfonts,amsmath}
\usepackage{epsfig}
\def\be{\begin{eqnarray}}
\def\ee{\end{eqnarray}}
\newcommand{\nn}{\nonumber}
\newcommand\para{\paragraph{}}
\newcommand{\ft}[2]{{\textstyle\frac{#1}{#2}}}
\newcommand{\eqn}[1]{(\ref{#1})}

\def\Dslash{\,\,{\raise.15ex\hbox{/}\mkern-12mu D}}
\def\Dbarslash{\,\,{\raise.15ex\hbox{/}\mkern-12mu {\bar D}}}
\def\delslash{\,\,{\raise.15ex\hbox{/}\mkern-9mu \partial}}
\def\delbarslash{\,\,{\raise.15ex\hbox{/}\mkern-9mu {\bar\partial}}}
\def\pslash{\,\,{\raise.15ex\hbox{/}\mkern-9mu p}}
\def\calDslash{\,\,{\raise.15ex\hbox{/}\mkern-12mu {\cal D}}}

\newcommand{\sign}{{\rm sign}}

\newcommand{\nin}{\,{\raise.15ex\hbox{/}\mkern-12mu \!\in}}

\newcommand{\CC}{{\mathbb C}}

\newcommand{\Tr}{{\rm Tr}}

\def\v{\varphi}

\def\implies{\Rightarrow}

\def\lae{\mathrel{\mathop{\smash{\lower .5 ex \hbox{$\stackrel<\sim$}}}}}
\def\lae{\mathrel{\mathop{\smash{\lower .5 ex \hbox{$\stackrel>\sim$}}}}}

\title{Dyonic Non-Abelian Vortices}
\author{Benjamin Collie \\
Department of Applied Mathematics and Theoretical Physics\\
University of Cambridge\\
Wilberforce Road, Cambridge CB3 0WA, UK\\
{\tt b.p.collie@damtp.cam.ac.uk}}

\abstract{We study three-dimensional Yang-Mills-Higgs theories with and without a Chern-Simons interaction.  We find that these theories admit a rich spectrum of vortex solitons carrying both a topological charge and a global flavour charge.  We further derive a low-energy description of the vortex dynamics from a gauged linear sigma model on the vortex worldline.}

\preprint{DAMTP-2008-76}

\begin{document}
\pagestyle{plain} \setcounter{page}{1}
\newcounter{bean}
\baselineskip16pt

\section{Introduction}

The discovery of vortices carrying genuine non-Abelian flux \cite{vib, vort12} has led to a series of studies aimed at developing an understanding of the dynamics of these objects.  Much of this work has focused on the role of vortex strings in $d=3+1$ dimensions, with the associated Higgs phases and the issue of dual confinement being of special interest.  It has also been suggested that non-Abelian vortices might have relevance for hot or dense QCD.  For reviews, see \cite{vort28, tasi, sy, QCD}.
\para

In this paper, we consider non-Abelian vortices in $d=2+1$ dimensional planar theories.  We are motivated by the possiblity that these solitons could be realised in condensed matter systems.  Studying theories both with and without a Chern-Simons interaction \cite{DJT}, we uncover a rich spectrum of vortices carrying both a topological charge and a global flavour charge.  We find that the dynamics of these dyonic objects depends intricately on the balance between the charges and the Chern-Simons coefficient.
\para

The plan of this paper is as follows.  In section \ref{3d} we introduce the model of interest, then find the Bogomolnyi bound on the energy of vortex configurations and show that the dyonic vortex mass is different in the cases with and without a Chern-Simons interaction.  In section \ref{1d}, we review how motion on the vortex moduli space is affected by massive matter and Chern-Simons interactions, and we rederive our vortex mass results from the perspective of a $d=0+1$ gauged linear sigma model.  Finally, in section \ref{Example}, we look at the implications of our results in the case of a single vortex in a $U(2)$ gauge theory.  We find that there is a variety of possible BPS and non-BPS solutions for the moduli space motion, including circular orbits and more exotic looping trajectories.

\section{Dyonic Vortices in $d=2+1$}
\label{3d}

We will begin by introducing the theory we are interested in, and highlighting some related models.  Then we will write down the energy functional for our theory and show that it is subject to a Bogomolnyi bound.  This bound gives the mass of a vortex.  We will see that the mass depends both on the topological charge and on a Noether charge associated with the global flavour symmetry of the theory.  The form of this dependence is different in the cases with and without a Chern-Simons interaction.
\para

We will work with a $U(N)$ Yang-Mills-Chern-Simons theory in $d=2+1$ dimensions, with a real adjoint scalar $\phi$ and $N_f$ fundamental scalars $q_i$ with real masses $m_i$, $i = 1,\dots,N_f$.  If the theory also has the appropriate fermions then it exhibits $\mathcal{N}=2$ supersymmetry (i.e. 4 supercharges), which determines the bosonic Lagrangian:
\be 
{\cal L} &=& -\frac{1}{2e^2}\Tr\,F_{\mu\nu}F^{\mu\nu} -
\frac{\kappa}{4\pi}\Tr\,\epsilon^{\mu\nu\rho}\left(A_\mu \partial_\nu
A_\rho - \frac{2i}{3}A_\mu A_\nu A_\rho\right) +
\frac{1}{e^2}\Tr\,({\cal D}_\mu\phi)^2 \nn\\ && + |{\cal D}_\mu
q_i|^2 - \sum_i q_i{}^{\!\dagger}(\phi-m_i)^2 q_i -
\frac{e^2}{4}\Tr\left(q_iq_i{}^{\!\dagger} -
\frac{\kappa\phi}{2\pi}-v^2\right)^2\label{2.1}.
\ee
For $N \geq 2$, the Chern-Simons coefficient $\kappa$ must be an integer so that the partition function is invariant under large gauge transformations.  For the Abelian $N=1$ theory, there is no such constraint.  Generically, the masses $m_i$ break the flavour symmetry of the model from $SU(N_f)$ to $U(1)^{N_f-1}$.
\para

This model is part of a large family of theories admitting vortex solutions.  We can obtain various related theories by taking limits of the above Lagrangian.
\begin{itemize}
\item 
When $\kappa=0$ and $m_i=0$, we have a Yang-Mills theory with massless fundamental scalar fields.  This theory admits non-Abelian vortices.  These were originally introduced in \cite{vib, vort12}, and have since been studied in some detail (for reviews, see \cite{vort28, tasi, sy}).
\item
When $\kappa=0$ but $m_i \neq 0$, we have a Yang-Mills theory with massive fundamental scalar fields.  The vortices of the $d=3+1$ version of this theory were discussed in \cite{4dGD, TongMono}.
\item
When $\kappa \neq 0$ but $m_i=0$, we have a Yang-Mills-Chern-Simons-Higgs theory like those studied in \cite{3dcs, Kao}.  This reduces to the Maxwell-Chern-Simons-Higgs theory of \cite{LLM, kimyeong} in the $N=1$ case.
\item
When $\kappa \neq 0$, $m_i=0$ and $e^2\rightarrow\infty$, the Yang-Mills term vanishes, and we can integrate out $\phi$ to get the Chern-Simons-Higgs theory with sixth order potential.  The Abelian version of this theory was first introduced in \cite{hkp, jlw, jw} and its non-Abelian generalisation has been studied more recently in \cite{klee, A&S}.
\end{itemize}
The review \cite{Dunne} covers many of the Chern-Simons models mentioned above.  Vortices in Yang-Mills-Chern-Simons theories with several Higgs fields and no fundamental matter fields have been previously studied in \cite{Schap1, Schap2, Schap3}.
\para

Our theory has vacua featuring different amounts of symmetry breaking.  We will work with the Higgs phase, which admits topologically stable vortices.  This phase exists only  when $N_f \geq N$; for simplicity, we set $N_f = N$ from here on.  In the Higgs phase, the vacuum expectation values of the fields are:
\be  \langle\phi\rangle = \mathrm{diag}\,(m_1, \ldots, m_N) \ \ \ \ ,\ \ \
\langle q_i^{\ a} \rangle = \delta_i^{\ a}\sqrt{v^2+\frac{\kappa}{2\pi}m_i}\label{HiggsPhase},\ee
where $a=1,\ldots,N$ is the colour index.   The flavour and gauge symmetries of the theory are each fully broken in this vacuum, but a subgroup of their product survives: $U(N)_{\rm gauge}\times U(1)^{N-1}_{\rm flavour}
\rightarrow U(1)^{N-1}_{\rm diag}$.  Note that we can always shift the vacuum expectation values to set $\sum_i m_i=0$.
\para

%In this section, we will write down the energy functional for the theory \eqn{2.1}, and show that it is subject to a Bogomolnyi bound.  This bound gives the mass of a vortex in the theory.  We will see that this mass depends both on the topological charge and on a Noether charge associated with the flavour symmetry of the theory.  The form of this dependence is different in the cases with and without a Chern-Simons interaction.
%\para
By a novel application of the Bogomolnyi procedure \cite{Bog, T6, LLM}, we will now find a lower bound on the energy of the field configurations of the theory.  Requiring that this bound should be saturated will give us a set of first order differential equations whose solutions are vortices.
\para

We first write down the energy functional $H$ of the theory.  This is the Noether charge associated with a combination of a time translation and a gauge transformation.  The gauge transformation is chosen so that the infinitesimal variation of any field under the combination is proportional to that field's covariant time derivative.
\be 
H &=& \int \mathrm{d}^2x \left[ \phantom{\left(\frac{\phi}{2}\right)^2}\!\!\!\!\!\!\!\!\!\!\!\!\!\!\! \frac{1}{e^2}\Tr\,\left(E_{\alpha}^2 + B^2 \right) + \frac{1}{e^2}\Tr\,\left(({\cal D}_0\phi)^2 + ({\cal D}_\alpha\phi)^2\right)\right. \nn
\\ &&\left. + |{\cal D}_0 q_i|^2 + |{\cal D}_\alpha q_i|^2+ \sum_i q_i{}^{\!\dagger}(\phi-m_i)^2 q_i +\frac{e^2}{4}\Tr\left(q_iq_i{}^{\!\dagger} -
\frac{\kappa\phi}{2\pi}-v^2\right)^2\right]\label{3.1}.
\ee
Note that we define $E_\alpha = F_{0\alpha}$ and $B = F_{12}$.  We now follow the usual Bogomolnyi procedure, and complete the square.
\be
\nonumber H &=&  \displaystyle\int \mathrm{d}^2 x \, \Bigg[ 
\lvert \mathcal{D}_{\pm} q_i \rvert^2 + \Tr\left( \tfrac{1}{e}B \pm \tfrac{1}{2}e \left( q_iq_i{}^{\!\dagger} -\tfrac{1}{2\pi}\kappa\phi - v^2 \right) \right)^2  
\\
&&\nonumber +\tfrac{1}{e^2}\Tr\left(\mathcal{D}_0\phi \right)^2 + \tfrac{1}{e^2} \Tr\left( E_\alpha \mp \mathcal{D}_\alpha\phi \right)^2  + \sum_i \lvert \mathcal{D}_0 q_i \mp i \left( \phi- m_i \right) q_i \rvert^2  
\\
&&\nonumber \mp \mathrm{Tr}\left[ 2\phi\left( -\tfrac{1}{4\pi}\kappa B + \tfrac{i}{2}\left[\left(\mathcal{D}_0 q_i \right) q_i{}^{\!\dagger}- q_i\left(\mathcal{D}_0 q_i \right)^{\dagger}\right] + \tfrac{1}{e^2}\mathcal{D}_\alpha E_\alpha + \tfrac{i}{e^2} \left[\mathcal{D}_0 \phi , \phi \right]\right)\right] 
\\
&&\pm \left( \mathrm{Tr}B \right)v^2 \pm i \sum_i\left[ \left( q_i{}^{\!\dagger}\left(\mathcal{D}_0 q_i \right) - \left(\mathcal{D}_0 q_i \right)^{\dagger}q_i \right)m_i\right]\Bigg].
\label{3.2}
\ee
In this expression we have used the notation $\mathcal{D}_{\pm} = \mathcal{D}_1 \pm i\mathcal{D}_2$.
\para

If $\kappa$ and $m_i$ are chosen to be zero then we may set $\phi \equiv 0$, so the third line in \eqn{3.2} disappears.  However, if either $\kappa$ or $m_i$ is nonzero and we have nontrivial fields $q_i$, $A_\alpha$, then the equation of motion for $\phi$ includes source terms, so we cannot set $\phi \equiv 0$.  Thus for the third line in \eqn{3.2} to disappear, we must invoke Gauss' law:
\be
-\frac{\kappa B}{4\pi} + \frac{i}{2}\left[\left(\mathcal{D}_0 q_i \right) q_i{}^{\!\dagger} - q_i\left(\mathcal{D}_0 q_i \right)^{\dagger}\right] + \frac{1}{e^2}\mathcal{D}_\alpha E_\alpha + \frac{i}{e^2} \left[\mathcal{D}_0 \phi , \phi \right] = 0. \label{3.4}
\ee
\para

Of the two terms in the last line of \eqn{3.2}, the first integrates to $\pm 2\pi kv^2$, where $k$ is the topological charge.  The integral of the second term can be written as $\pm \sum_i Q_i m_i$, where we define
\be
Q_i \equiv i \displaystyle\int \mathrm{d}^2 x \left( q_i{}^{\!\dagger}\left(\mathcal{D}_0 q_i \right) - \left(\mathcal{D}_0 q_i \right)^{\dagger}q_i \right).  \label{Q_i}
\ee
The $Q_i$ are conserved Noether charges associated with the $U(1)^{N-1}$ flavour symmetry of the theory.  If we take the trace of Gauss' law \eqn{3.4} and then integrate over space, we find that $\sum_i Q_i = \kappa k$.
\para

All of the terms in the first two lines of \eqn{3.2} are non-negative squares, so we see that the energy $H$ of our configuration is subject to a Bogomolnyi bound, which we interpret as the vortex mass.
\be
H \geq \left| 2\pi k v^2 + \sum_i Q_i m_i \right| \equiv M_{\rm vortex}.  \label{3.8}
\ee
This result is remarkable because it means that sometimes increasing the value of $|\sum_i Q_i m_i|$ may decrease the vortex mass.  This is an unusual property: typically, increasing the size of a charge on a soliton will increase the soliton's mass.
\para

To saturate the bound \eqn{3.8}, a configuration must obey Gauss' law \eqn{3.4} and solve the first order Bogomolnyi equations that arise from setting the squared terms to zero in \eqn{3.2}.  Given a configuration that satisfies these constraints, it is straightforward to show that the quantity $2\pi k v^2 + \sum_i Q_i m_i$ is strictly positive if $k>0$ and strictly negative if $k<0$.  For the case $k=1$, we will show in the next section that the minimum possible value of $M_{\rm vortex}$ for given $m_i$ is $2\pi v^2 + m_p \kappa$, where $m_p$ is the most negative of the $m_i$ values if $\kappa>0$ or the most positive of the $m_i$ values if $\kappa<0$.
\para

In the special case $\kappa=0$, the Bogomolnyi process works slightly differently.  The absence of the $\kappa B$ term in Gauss' law means that in \eqn{3.2} we can make one sign choice for the first line and the topological charge term, and a separate choice for the second and third lines and the flavour charge term.  As a result, we have a stricter Bogomolnyi bound:
\be
H \geq \left| 2\pi k v^2 \right| + \left| \sum_i Q_i m_i \right| \equiv M_{\rm vortex}.  \label{3.11}
\ee
Once again, the bound is saturated if and only if Gauss' law and the Bogomolnyi equations hold.
\para  

We call the vortices in this theory dyonic because $M_{\rm vortex}$ has contributions from two types of charge, the topological charge $k$ and the Noether charge $\sum_i Q_i m_i$.  The significance of the difference between the bounds \eqn{3.8} and \eqn{3.11} is that increasing $\left| \sum_i Q_i m_i \right|$ always increases the vortex mass in the $\kappa=0$ case, but may either increase or decrease the vortex mass in the $\kappa \neq 0$ case.
\para

The dyonic vortices we have considered in $d=2+1$ dimensions are related to several other kinds of dyonic solitons in different numbers of dimensions.  These include dyonic instantons in $d=4+1$ \cite{Y16}, the original dyonic monopoles in $d=3+1$ \cite{T99, T98, WittenDyons}, semilocal vortices in $d=2+1$ \cite{Y15}\footnote{Semilocal vortices can arise in Abelian gauge theories with multiple flavours of matter, and are related to sigma model Q-lumps.  The moduli space of semilocal vortices suffers from non-normalizable zero modes corresponding to the scale parameters of the vortices \cite{Y20,Y21}.  Our non-Abelian model avoids this problem because it contains equal numbers of colours and flavours: it hence has no vortex scale parameters, so the vortex moduli space is free of non-normalizable zero modes.} and the dyonic domain walls known as Q-kinks in $d=1+1$ \cite{Y18}.  A wide variety of soliton states, including vortices and their dyonic extensions, were considered in \cite{Nitta1, Nitta2, Nitta3, Nitta4}.
\para

It is interesting to compare the dyonic vortex masses we have calculated with the masses of other types of dyons.  For dyonic monopoles and Q-kinks, the mass can be written schematically as $M=\sqrt{Q_a^2 + Q_b^2}$, where $Q_a$ and $Q_b$ are the two types of charge that make these solitons dyonic \cite{Y17,Y18}.  We can therefore regard dyonic monopoles and Q-kinks as bound states, each containing two distinct objects with masses $Q_a$ and $Q_b$.  For dyonic instantons and semilocal vortices, the mass is schematically closer to $M=Q_a + Q_b$ \cite{Y16, Y15}, so it seems these objects are only threshold bound states.
\para

The dyonic vortices in our study, meanwhile, can be regarded as true bound states of two different types of object.  The first object is a vortex with a topological charge but no flavour charges; this has mass $2\pi |k| v^2$.  The second object has flavour charges but no topological charge, so it is made up of excitations of the squark fields $q_i^a$.  Excitations of the $q_i^a$ about the vacuum have masses $\sqrt{e^2\langle q_i^a \rangle^2/2 +(m_a-m_i)^2}$.  The dyonic vortex mass $M_{\rm vortex}$ is smaller than the sum of the masses of the constituent objects, since the $\langle q_i^a \rangle$ part of the squark excitation masses does not contribute to $M_{\rm vortex}$.  Hence the dyonic vortex is a classically stable bound state.

\section{The Dyonic Vortex Worldline Theory}
\label{1d}

The moduli space approximation is a useful tool for understanding the low-energy dynamics of vortices \cite{manton}.  For the theory given by \eqn{2.1}, with $\kappa$ and $m_i$ set to zero, slow motion on the vortex moduli space is described by a nonlinear sigma model.  In this section, we will review how this motion is modified when we reintroduce nonzero $m_i$ and $\kappa$.  We will then reformulate our description of the motion as a gauged linear sigma model, and see how from this perspective we can rederive the results of the preceding section by applying the Bogomolnyi procedure to the $d=0+1$ energy functional.
\para

In the theory \eqn{2.1} with $m_i=0$ and $\kappa=0$, the set of static physical solutions of the Bogomolnyi equations with topological charge $k$ forms a moduli space $\mathcal{M}$.  This has collective co-ordinates $X^p$ and K\"ahler metric $g_{pq}$, where $p,q=1,\ldots,2N|k|$.  Low-energy vortex dynamics is described by geodesic motion on $\mathcal{M}$:
\be 
L = \ft12
g_{pq}(X)\dot{X}^p\dot{X}^q. \label{4.1}
\ee
The metric has an $SU(N)$ isometry arising from the unbroken $SU(N)_{\rm diag}$ remnant of the gauge and flavour symmetries of the theory.  We can find a set of $N$ Killing vectors $\mathcal{K}_i$ associated with the isometry such that $\sum_i \mathcal{K}_i = 0$.
\para

If we allow $m_i \neq 0$ but keep $\kappa = 0$, then a potential is induced on the moduli space.  This potential can be written in terms of the Killing vectors $\mathcal{K}_i$ \cite{4dGD, TongMono, Y11}:
\be
V_m = \sum_{i,j=1}^N \left( m_i \mathcal{K}_i^p \right) \left( m_j \mathcal{K}_j^q \right)^\dagger g_{pq}. \label{4.V}
\ee
If we instead allow $\kappa \neq 0$ but keep $m_i = 0$, then a magnetic field $\mathcal{F} \in \Omega^2\left(\mathcal{M}\right)$ is induced on the moduli space.  Writing this locally as $\mathcal{F} = d\mathcal{A}$ and working to leading order in $\kappa e/v$, we find the dynamics on $\mathcal{M}$ is modified by an extra term in the Lagrangian \cite{3dcs, kimyeong}:
\be
L_\kappa =  -\kappa {\cal A}_p(X)\dot{X}^p. \label{4.K}
\ee
If we allow both $m_i \neq 0$ and $\kappa \neq 0$, then the Lagrangian for motion on $\mathcal{M}$ includes both the Chern-Simons term \eqn{4.K} and a potential similar to \eqn{4.V}.  We might expect that the potential \eqn{4.V} is modified by terms of order $m_i\kappa$; we have not calculated this from the nonlinear sigma model, but in the next subsection we will check that it is true, at least for $k=1$, by using a gauged linear sigma model.
\para

\subsection*{The Gauged Linear Sigma Model For One Vortex}

For an alternative description of motion on the one-vortex moduli space, and to clarify what happens when $\kappa$ and $m_i$ are both nonzero, we now assemble a standard $d=0+1$ gauged linear sigma model, with a $U(1)$ gauge field and $\mathcal{N}=(0,2)$ supersymmetry.  This `worldline theory' for a non-Abelian vortex is straightforward to solve, but it is more interesting than the corresponding theory for an Abelian vortex because non-Abelian vortices have an internal orientation as well as a spatial position.  If we ignore the spatial position of our non-Abelian vortex and set $m_i=0$, then the moduli space of vortex configurations is $\mathcal{M}=\mathbb{CP}^{N-1}$, with homogeneous co-ordinates $\v_i \in \CC$, $i = 1, \ldots, N$.  These co-ordinates give $N$ fundamental scalar fields in the worldline theory, and roughly speaking they correspond to the orientation of the vortex in the $U(N)$ gauge group of the parent theory.  The fields $\v_i$ are subject to the $D$-term constraint
\be
D \equiv \sum_{i=1}^N \v_i \v_i{}^{\!\dagger} -r = 0, \label{D-term}
\ee
where $r$ is a constant which is determined to be $2\pi/e^2$ \cite{vib, vort12, vort39}.  In the case $m_i=0$, $\kappa=0$, the Lagrangian \eqn{4.1} is $L = \left| \mathcal{D}_t \v_i \right|^2$, where $\mathcal{D}_t \v_i = \partial_t \v_i - iA\v_i$ and $A$ is the gauge field for the $U(1)$ worldline gauge symmetry.  The scalars $\v_i$ enjoy an $SU(N)$ flavour symmetry corresponding to the $SU(N)$ isometry of the metric on $\mathcal{M}$.  
\para

If we introduce nonzero masses $m_i$ for the $q_i$ of the parent theory, then the gauged linear sigma model is deformed by the introduction of masses $m_i$ for the $\v_i$ \cite{4dGD}:
\be
L = \left| \mathcal{D}_t \v_i \right|^2  - \sum_i \v_i{}^{\!\dagger}\left( \sigma - m_i \right)^2 \v_i.  \label{4.m}
\ee
Here, $\sigma$ is an adjoint scalar field in the $\mathcal{N} = (0,2)$ gauge multiplet.  The deformation \eqn{4.m} is consistent with the $\mathcal{N} = (0,2)$ supersymmetry of the theory, but for generic $m_i$ it breaks the flavour symmetry from $SU(N)$ to $U(1)^{N-1}$.  When we integrate out $\sigma$ and $A$ in \eqn{4.m}, we recover the moduli space potential \eqn{4.V}.
%
%To see this, we begin by rewriting the deformation in terms of the metric $g_{pq}$, decomposing the indices as $p=(I,\alpha)$, $q=(J,\beta)$, where $I,J=1,\ldots,N$.
%\be \nn
%-L_m = \sum_{I,J} \left[ \left(\sigma^{\alpha\gamma}\v_I^\gamma\right)^\dagger \left(\sigma^{\beta\delta}\v_J^\delta\right) - 2 \left(\sigma^{\alpha\gamma}\v_I^\gamma\right)^\dagger \left(m_J\v_J^\beta\right) + \left(m_I\v_I^\alpha\right)^\dagger \left(m_J\v_J^\beta\right) \right] g_{(I,\alpha)(J,\beta)}.
%\ee
%Next, we use the constraint \eqn{D-term} and the field equation for $\sigma$ to write
%\be \nn
%-L_m &=& \sum_{i,j=1}^N m_i m_j \v_i^\alpha \v_j^{\beta\dagger} \left( \delta_{Ii}\delta_{Jj} - \tfrac{1}{r} \v_I^\gamma \v_i^{\gamma\dagger}\delta_{Jj} - \tfrac{1}{r} \v_J^{\gamma\dagger} \v_j^\gamma\delta_{Ii} + \tfrac{1}{r^2} \v_I^\gamma \v_i^{\gamma\dagger} \v_J^{\delta\dagger} \v_j^\delta \right) g_{(I,\alpha)(J,\beta)}.
%\ee
%This reduces to the potential \eqn{4.V} as required when we apply \eqn{N.1}.
\para

If we now introduce a Chern-Simons interaction into the parent theory, then a corresponding term is induced in the gauged linear sigma model \cite{3dcs}:
\be
L = \left| \mathcal{D}_t \v_i \right|^2 - \sum_i \v_i{}^{\!\dagger}\left( \sigma - m_i \right)^2 \v_i -\kappa\left(A+\sigma\right).  \label{4.2Ab}
\ee
Note that we can again integrate out $\sigma$ and $A$ to find the potential on $\mathcal{M}$:
\be
V = \frac{1}{r}\sum_{i<j} (m_i-m_j)^2 |\v_i|^2|\v_j|^2 + \frac{\kappa}{r} \sum_i m_i |\v_i|^2.  \label{3.7}
\ee
The first term is independent of $\kappa$, and matches the potential in \cite{4dGD}.  The second term is the promised order $m_i\kappa$ modification.  Note that the first term is non-negative, so $V$ can be minimised (subject to the constraint \eqn{D-term}) by choosing $p \in {1,\ldots,N}$ such that $m_p\kappa \leq m_i\kappa$ $\forall i$, then setting
\be
|\v_i| = \left\{ \begin{array}{cl} r \qquad & \mathrm{if}\qquad i=p \\ 0 \qquad & \mathrm{otherwise.} \end{array} \right. \label{MinMass}
\ee
\para

From the Lagrangian \eqn{4.2Ab}, we can compute Gauss' law,
\be
&&i \sum_{i=1}^N \left( \left( \mathcal{D}_t \v_i \right) \v_i{}^{\!\dagger} - \v_i \left( \mathcal{D}_t \v_i \right)^\dagger \right) = \kappa \label{4.3} \\
&&\implies \sum_i \mathcal{Q}_i = \kappa  \ \ \ , \ \ \ {\rm where} \ \ \ \mathcal{Q}_i \equiv i \left( \left( \mathcal{D}_t \v_i \right)\v_i{}^{\!\dagger}  - \v_i\left( \mathcal{D}_t \v_i \right)^\dagger  \right).  \label{4.C}
\ee
The $\mathcal{Q}_i$ are conserved Noether charges corresponding to the $U(1)^{N-1}$ flavour symmetry of the worldline theory.  As we will see soon, they are related to the Noether charges $Q_i$ \eqn{Q_i} of the parent theory.
\para

\subsection*{The Dyonic Vortex Mass}

We are now ready to apply the Bogomolnyi procedure to rederive the results of section \ref{3d}.  As in the $d=2+1$ case, we can find the energy functional for the theory as the Noether charge associated with a combined time translation and gauge transformation.
\be
H_{1d} = \left| \mathcal{D}_t \v_i \right|^2 + \sum_i \v_i{}^{\!\dagger}\left( \sigma - m_i \right)^2 \v_i  + \kappa\sigma.  \label{08-07-3.10}
\ee
When we complete the square, we get
\be \nn
H_{1d} &=& \left| \mathcal{D}_t \v_i - i (\sigma - m_i) \v_i \right|^2  \\ 
&&+ \Tr \left[ 
\sigma \left( \kappa - i\sum_i \left( \left( \mathcal{D}_t \v_i \right) \v_i{}^{\!\dagger} - \v_i \left( \mathcal{D}_t \v_i \right)^\dagger\right) 
\right)  \right] +\sum_i \mathcal{Q}_i m_i. \label{D1.4.5}
\ee
Imposing Gauss' law \eqn{4.3} gives the lower bound $H_{1d} \geq \sum_i \mathcal{Q}_i m_i$.  Note that if $m_i=0$, then the vortex has mass $2\pi v^2$.  The energy $H_{1d}$ gives corrections to this mass which arise from the fields on the worldline, so we can write
\be 
M_{\rm vortex} = 2\pi v^2 + \sum_i \mathcal{Q}_i m_i. \label{4.A}
\ee
This formula is remarkable because it shows from the worldline perspective that appropriately chosen values for the charges $\mathcal{Q}_i$ can give a negative contribution to the mass of the vortex.  This is an unusual property for a soliton, but it matches what we found in the parent theory in section \ref{3d}.
\para

By eliminating the auxiliary fields $\sigma$ and $A$ from \eqn{08-07-3.10}, we can see that the minimal possible value of $M_{\rm vortex}$ for given $m_i$ is achieved by setting $\dot{\v}_i=0$ and then choosing $\v_i$ to satisfy \eqn{MinMass}, so that $V$ is minimised.  We then have $M_{\rm vortex}=2\pi v^2 +m_p\kappa$, where $m_p$ is the most negative of the $m_i$ values if $\kappa>0$ or the most positive of the $m_i$ values if $\kappa<0$.  This minimal value of $M_{\rm vortex}$ is guaranteed to be strictly positive by the fact that the vacuum expectation value $\langle q_p^p \rangle$ given in \eqn{HiggsPhase} must be strictly positive to permit the winding of the vortex to lie in the $p$th flavour, as demanded by \eqn{MinMass}.
\para

In the special case $\kappa=0$, the Bogomolnyi process works slightly differently, just as in three dimensions.  In particular, we have a choice of signs when we complete the square:
\be \nn
H_{1d} &=& \left| \mathcal{D}_t \v_i \pm i (\sigma - m_i) \v_i \right|^2 \\ 
&&\pm \Tr \left[ 
\sigma \left( i\sum_i \left( \left( \mathcal{D}_t \v_i \right) \v_i{}^{\!\dagger} - \v_i \left( \mathcal{D}_t \v_i \right)^\dagger\right) 
\right)\right]  \mp \sum_i \mathcal{Q}_i m_i.  \label{D1.4.8}
\ee
This gives us a stricter bound on the energy once we apply Gauss' law.
\be 
H_{1d} \geq \left| \sum_i \mathcal{Q}_i m_i \right|
\implies M_{\rm vortex} = \left| 2\pi k v^2 \right| + \left| \sum_i \mathcal{Q}_i m_i \right|. \label{4.B}
\ee
\para

Our results \eqn{4.A}, \eqn{4.B} match the $k=1$ versions of the corresponding results \eqn{3.8}, \eqn{3.11} from the parent theory if we make the identification $Q_i = \mathcal{Q}_i$.  Note that under this identification, there is agreement between the constraints $\sum_i Q_i = \kappa k$ and $\sum_i \mathcal{Q}_i = \kappa$, and the right hand side of \eqn{4.A} is necessarily positive.  Just as we saw in section \ref{3d}, increasing $\left| \sum_i \mathcal{Q}_i m_i \right|$ always increases the vortex mass in the $\kappa=0$ case, but may either increase or decrease the vortex mass in the $\kappa \neq 0 $ case.
\para

If we are interested in a multi-vortex system, it is again possible to provide a gauged linear sigma model for the dynamics, this time by using a brane construction \cite{vib}.  However, the metric induced on the moduli space when we quotient the gauged linear sigma model by the action of the gauge group no longer matches the metric $g_{pq}$ of the nonlinear sigma model.  Consequently, the $k$-vortex gauged linear sigma model does not give a strictly accurate description of $k$-vortex moduli space dynamics, although it may still be useful for finding qualitative properties of the motion.

\section{Example: A Single $U(2)$ Vortex}
\label{Example}

In this section, we will apply our results to the case of a single vortex in a $U(2)$ Yang-Mills-Chern-Simons theory including two flavours of matter with masses $m$ and $-m$.  We will obtain the effective potential on the moduli space of vortices, and use it to study some features of the dynamics.  In particular, we will find that the gauged linear sigma model perspective makes it straightforward to find BPS and non-BPS solutions corresponding to several different types of moduli space motion.
\para

If $m=0$, the moduli space $\mathcal{M}$ is ${\bf S}^2\cong{\bf CP}^1$.  %  Co-ordinates on this space are given by
%
%\be \varphi_1 = \sqrt{r} e^{i\psi-i\phi/2}\cos(\theta/2) \ \ \ ,\
%\% \ \varphi_2 = \sqrt{r}
%e^{i\psi+i\phi/2}\sin(\theta/2)\label{coords},  \label{5.1}\ee
%
%where the co-ordinate $\psi$ is associated with the $U(1)$ gauge symmetry of the worldline theory; $\psi \in[0,2\pi)$, $\phi \in [0,2\pi)$ and $\theta \in [0, \pi]$.  
When we switch on the mass $m$, a potential given by \eqn{3.7} is induced on $\mathcal{M}$.  Using co-ordinates $\theta \in [0, \pi]$ and $\phi \in [0,2\pi)$ on ${\bf S}^2$, we have
\be 
L= \frac{r}{4}\left[\dot{\theta}{}^2+
\sin^2\theta\,\dot{\phi}{}^2\right] +
\frac{\kappa}{2}\left( \cos\theta - 1 \right)\dot{\phi}  - rm^2\sin^2\theta - \kappa m \cos\theta .
\label{5.2}
\ee
The first term describes a standard sigma model on ${\bf S}^2$.  The second term is the Dirac monopole connection, as was explained in \cite{3dcs}, and the final two terms give the potential induced by $m$.  Changing the sign of $m$ corresponds to changing which pole we identify as north.  Changing the sign of $\kappa$ corresponds to performing a parity transformation $\theta \rightarrow \pi-\theta$.  In what follows, we choose both $m$ and $\kappa$ to be non-negative without loss of generality.
\para

The theory described by \eqn{5.2} has a conserved angular momentum:
\be
J = r \dot{\phi} \sin^2 \theta + \kappa \cos\theta = \mathcal{Q}_1 - \mathcal{Q}_2. \label{5.J}
\ee
%The theory described by \eqn{5.2} has a conserved angular momentum $J$ associated with the co-ordinate $\phi$, 
There is also a conserved Hamiltonian which we can write as 
\be
H = \tfrac{r}{4} \dot{\theta}^2 + V_{\rm eff}(\theta), \ \ \ {\rm where} \ \ \ V_{\rm eff}(\theta) = \frac{(J-\kappa\cos\theta)^2}{4r\sin^2\theta} + rm^2\sin^2\theta + m\kappa\cos\theta.
\ee
The shape of the effective potential $V_{\rm eff}$ determines the dynamics of the system.  %Introducing the quantities $\hat{\kappa}=\kappa/(2mr)$, $\hat{J}=J/(2mr)$, and w
Writing $c=\cos\theta$, $s=\sin\theta$, we find
\be
V_{\rm eff}'(\theta) = 2rm^2s^{-3} \left(c^2 - \frac{\kappa}{2mr}c + \frac{J}{2mr}-1 \right) \left(c^3 - c\left(\frac{J}{2mr}+1\right) + \frac{\kappa}{2mr} \right).
\ee

\subsection*{The Shape Of $V_{\rm eff}$ When $\kappa=0$ And $m \neq 0$}

When $J=0$, $V_{\rm eff}$ has a maximum at $\theta=\pi/2$ and two minima at $\theta=0,\pi$.  When $|J|>0$, $V_{\rm eff} \rightarrow \infty$ as $\theta \rightarrow 0$ or $\pi$.  For $0<|J|<2mr$, $V_{\rm eff}$ has a maximum at $\theta = \pi/2$ and two minima at the values of $\theta$ given by $\sin\theta = \sqrt{|J|/(2mr)}$, as shown in figure 1.  For $|J|\geq 2mr$, $V_{\rm eff}$ has a single minimum at $\theta=\pi/2$.
\para

The energy of a solution sitting at a minimum of $V_{\rm eff}$ is given by 
\be
H_{1d} = \left\{ 
\begin{array}{rlcr}
2rm^2 & |J/(2mr)| & \qquad {\rm for} & |J| \leq 2mr \\
rm^2 & [(J/(2mr))^2+1] & \qquad {\rm for} & |J| > 2mr.
\end{array} \right.
\ee
\EPSFIGURE{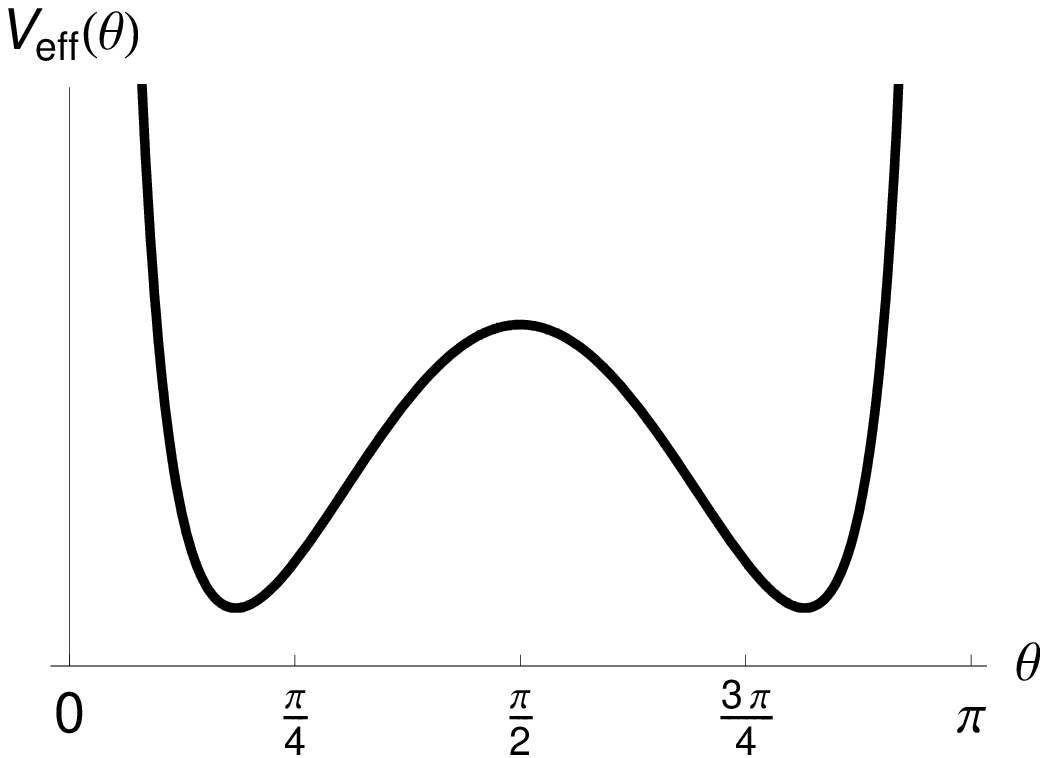,width=7cm}{A sketch of the effective potential when $\kappa=0$ and $0<| J|<2mr$.} 
To understand this, we turn to the Bogomolnyi equations that arise from \eqn{D1.4.8}.  These are satisfied by setting $\dot{\theta}=0$ along with either $\theta=0$, $\theta=\pi$ or $\dot{\phi}=2m\,\sign (mJ)$.  For $|J| \leq 2mr$, solutions to the Bogomolnyi equations are precisely the configurations that sit at minima of $V_{\rm eff}$; these saturate the Bogomolnyi bound 
$
H_{1d} \geq  \left| \sum_i \mathcal{Q}_i m_i \right| = |mJ|
$.
For $|J|> 2mr$, the Bogomolnyi equations cannot be solved, and the energy exceeds the Bogomolnyi bound even at the minimum of $V_{\rm eff}$.

\subsection*{The Shape Of $V_{\rm eff}$ When $\kappa > 0$}

For nonzero $\kappa$, $V_{\rm eff}$ can take on a rich variety of different shapes.  We find different patterns of behaviour in the cases $0<\kappa<mr/2$, $mr/2<\kappa<4mr$ and $4mr<\kappa$, and in this section we will consider each of these cases in turn.  
\para

In every case, we have $ V_{\rm eff} \rightarrow \infty $ as $\theta \rightarrow 0$ or $\pi$, except where stated otherwise.  The Bogomolnyi equations arising from \eqn{D1.4.5} are solved by setting $\dot{\theta}=0$ and either $\theta=0$, $\theta=\pi$ or $\dot{\phi}=2m$; note that by \eqn{5.J}, $\dot{\phi}$ is a function of $\theta$.  To satisfy $\dot{\theta}=0$, a configuration must sit at an extremum of $V_{\rm eff}$.  We will refer to a minimum of $V_{\rm eff}$ as `BPS' if it is located at a value of $\theta$ such that $\theta=0$, $\theta=\pi$ or $\dot{\phi}=2m$.  Otherwise, we will say the minimum is `non-BPS'.  At BPS minima, the Bogomolnyi bound is saturated, so $V_{\rm eff}=mJ$.  At non-BPS minima, $V_{\rm eff}>mJ$.  We will see that minima of $V_{\rm eff}$ often occur at $\theta = f_\pm$, where we define
\be \nn
f_\pm(\kappa,J) = \cos^{-1}\left(\frac{1}{2}\left(\frac{\kappa}{2mr} \pm \sqrt{ \left(\frac{\kappa}{2mr}\right)^2 - 4\left( \frac{J}{2mr}-1 \right) } \right)\right).
\ee
\para
\EPSFIGURE{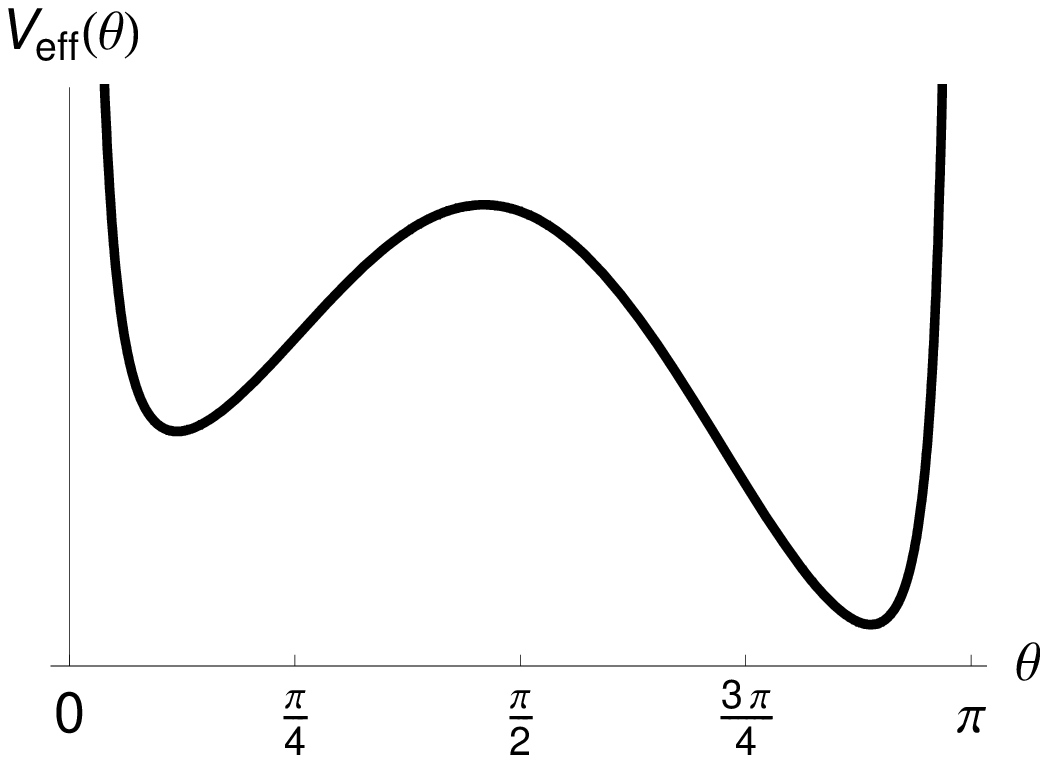,width=7cm}{A sketch of the effective potential when $0<\kappa<mr/2$ and $|J| < \kappa$.}

We first describe the shape of $V_{\rm eff}$ in the case $0<\kappa<mr/2$ for different values of $J$.  When $|J|<\kappa$, $V_{\rm eff}$ has a non-BPS minimum and a maximum at values of $\theta$ less than $\pi/2$, and a BPS minimum at $\theta = f_-$.  A sketch is given in figure 2.  When $\kappa \leq J < 2mr+\kappa^2/(8mr)$, $V_{\rm eff}$ has two BPS minima at $\theta=f_\pm$, with a maximum in between.  A sketch of $V_{\rm eff}$ would then look like figure 1, but with the positions of the extrema shifted sideways; an exception is the special case $J=\kappa$, when one of the minima is actually at $\theta=0$.  When $J >  2mr+\kappa^2/(8mr)$, $V_{\rm eff}$ has a single non-BPS minimum at a value of $\theta$ in $(0,\pi/2)$.  
\para

Now suppose we decrease $J$ until it is negative.  When $J= -\kappa$, $V_{\rm eff}$ has a BPS minimum at $\theta=\pi$, and a non-BPS minimum and a maximum at values of $\theta$ in $(0,\pi/2)$.  For $2mr(3[\kappa/(4mr)]^{2/3} -1)< J < -\kappa$, $V_{\rm eff}$ has a non-BPS minimum at a value of $\theta$ in $(\pi/2,\pi)$, and a non-BPS minimum and a maximum at values of $\theta$ in $(0,\pi/2)$.  A sketch of $V_{\rm eff}$ would look similar to figure 2.  Finally, when $J<2mr(3[\kappa/(4mr)]^{2/3} -1)$, $V_{\rm eff}$ has a single non-BPS minimum at a value of $\theta$ in $(\pi/2,\pi)$.  This concludes our description of possible shapes of $V_{\rm eff}$ when $0<\kappa<mr/2$.
\para

The pattern of possible shapes is slightly different if we choose $mr/2<\kappa<4mr$ because now $-\kappa < 2mr(3[\kappa/(4mr)]^{2/3} -1)$.  For $-\kappa \leq J \leq 2mr(3[\kappa/(4mr)]^{2/3} -1)$, $V_{\rm eff}$ has a single BPS minimum at $\theta=f_-$; note that $f_-=\pi$ for $J=-\kappa$.  For $J<-\kappa$ or $J > 2mr(3[\kappa/(4mr)]^{2/3} -1)$, the behaviour matches what we had in the case $0<\kappa<mr/2$ for $J < 2mr(3[\kappa/(4mr)]^{2/3} -1)$ or $J>-\kappa$ respectively.
\para
\EPSFIGURE{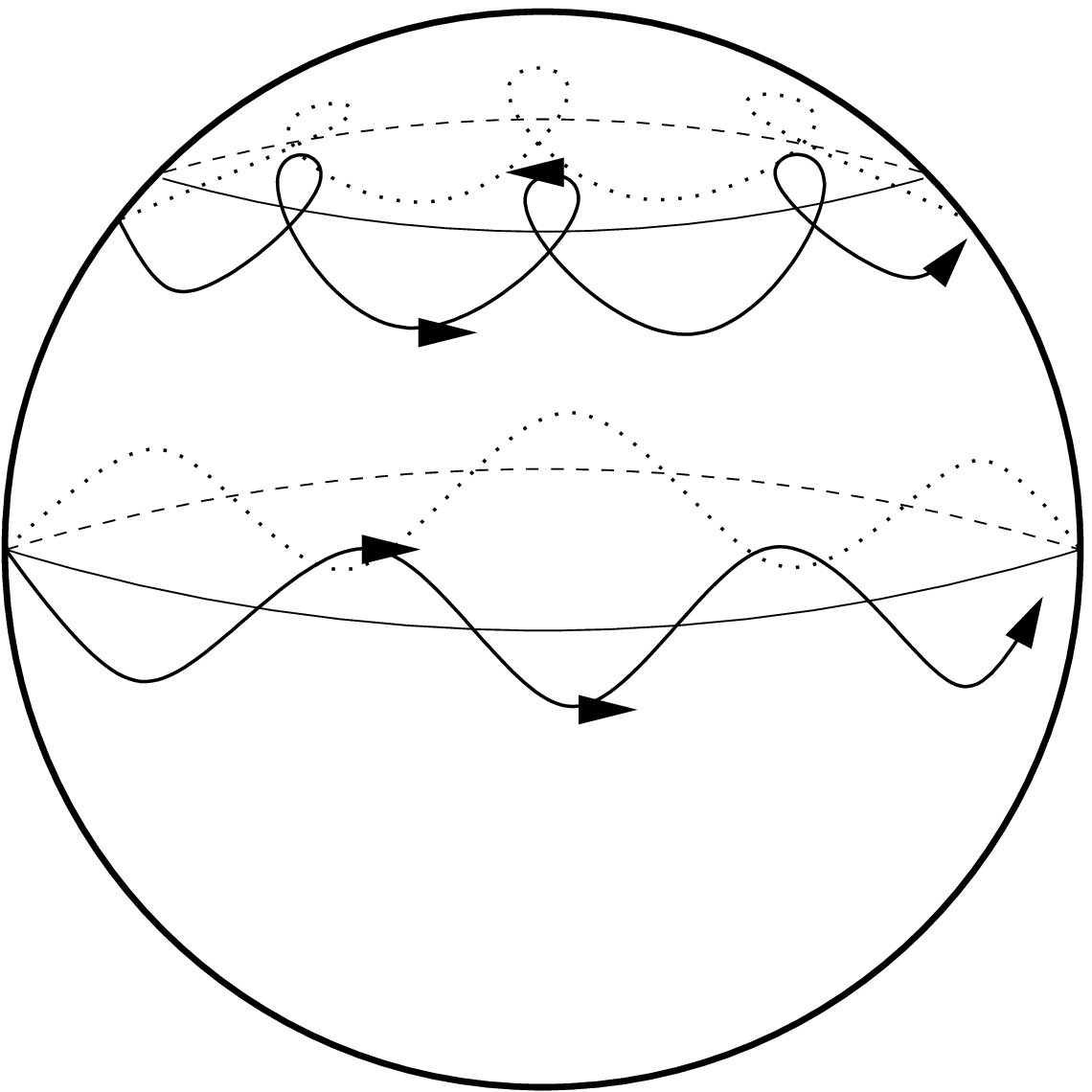,width=6.5cm}{Possible moduli space paths with nonzero $\kappa$ and $m$.}
In the case $\kappa>4mr$, $V_{\rm eff}$ always has a single minimum.  For $|J|\leq\kappa$, the minimum is BPS and is located at $\theta=f_-$.  In particular, the minimum is at $\theta=0$ when $J=\kappa$ and at $\theta=\pi$ when $J=-\kappa$.  For $J<-\kappa$, the minimum is non-BPS and is at a value of $\theta$ in $(0, \pi/2)$; for $J>\kappa$, the minimum is non-BPS and is at a value of $\theta$ in $(\pi/2,\pi)$.  Note that the results of this paragraph still apply in the special case $m=0$ if we redefine $f_-=\cos^{-1}(J/\kappa)$.

\subsection*{Possible Types Of Motion}

Each minimum of $V_{\rm eff}$ gives a stable solution of the equations of motion of the system.  Usually, such a solution is a circular orbit around ${\bf S}^2$ with fixed nonzero $\dot{\phi}$ at a fixed value of $\theta$.  However, if the minimum occurs at $\theta=0$ or $\pi$, then it gives a static solution sitting at a pole of ${\bf S}^2$.  Also, for $m=0$ and $|J|<\kappa$, the BPS minimum of $V_{\rm eff}$ corresponds to a static solution with $\dot{\phi}=0$ sitting at a value of $\theta$ in $(0,\pi)$: at this value of $\theta$, the effects of $J$ and $\kappa$ effectively cancel one another out.
\para

If we give the system an energy greater than that required to sit at a given minimum of $V_{\rm eff}$, then $\theta$ oscillates around that minimum and the motion on ${\bf S}^2$ can take two distinct forms, as shown in figure 3.  For $|J|>\kappa$, the sign of $\dot{\phi}$ always matches that of $J$.  We then have motion similar to the lower, wave-like path in figure 3.  Things are more interesting for $|J|<\kappa$.  Then 
\be \nn
\sign\left(\dot{\phi}\right) = \sign\left( \theta - \cos^{-1}\left( \frac{J}{\kappa} \right)\right).
\ee
If the minimum of $V_{\rm eff}$ that we are oscillating around is at the critical value $\theta = \cos^{-1} (J/\kappa)$ then we always have motion similar to the upper, looping path in figure 3.  If the minimum of $V_{\rm eff}$ that we are oscillating around is not at the critical value of $\theta$, then we can have a wave-like path or a looping path depending on whether there is enough energy for $\theta$ to cross the critical value as it oscillates.

%
%\DOUBLEFIGURE{DyonicVeff3.eps,width=7cm}{Sphere2.eps,width=6.5cm}{The effective potential when $\kappa=0$ and $0<| \hat{J} |<1$.}{A small $\hat{\kappa}$ shifts both circular orbits towards the same pole.} 
%

%For $\left|\hat{\kappa}\right| \ll 1$, $V_{\rm eff}$ is deformed only slightly from the $\hat{\kappa}=0$ case of figure 1.  If we take $0<|\hat{J}|<1$, then the two minima of $V_{\rm eff}$ which were located at $\theta_\pm$ in the $\hat{\kappa}=0$ case are shifted to $\theta_\pm + \epsilon_\pm$, where $\epsilon_\pm = O\left(\hat{\kappa}\right)$.  A straightforward calculation shows that $\epsilon_+=\epsilon_-$, so the direction of the shift does not depend on which minimum of $V_{\rm eff}$ we started off at.  Both of the stable circular orbits around ${\bf S}^2$ are shifted towards the same pole, so the radius of one of the orbits gets a little larger and the radius of the other gets a little narrower, as shown in Figure 2.  This happens because the Chern-Simons term in \eqn{5.2} behaves like a magnetic field, exerting a force in a direction determined by the sign of $\kappa / J$.
%\para

\setcounter{section}{0} \setcounter{equation}{0}
\renewcommand{\thesection}{\Alph{section}}

\section*{Acknowledgement}
I would like to thank Nick Dorey, Nick Manton and David Tong
for many useful discussions. I am supported by an STFC studentship.

\end{document}